\documentclass[iop,revtex4,twocolumn,nofootinbib,apj]{emulateapj}

\usepackage{amsfonts}
\usepackage{amsmath}
\usepackage{amssymb}
\usepackage{upgreek}
\usepackage{bm}
\usepackage{dcolumn}
\usepackage{epsfig}
\usepackage{graphicx}
\usepackage{placeins}
\usepackage{dsfont}
\usepackage{comment}
\usepackage{mathtools}
\usepackage{wrapfig}
\usepackage{lscape}
\usepackage{rotating}
\usepackage{epstopdf}
\usepackage{courier}
\usepackage{color}

\shorttitle{Constraining BBH Formation Models with GW Observations}
\shortauthors{Zevin et al. 2017}

\begin{document}

\title{Constraining Formation Models of Binary Black Holes\\ with Gravitational-Wave Observations}

\author{Michael Zevin$^1$}
\author{Chris Pankow$^1$}
\author{Carl L.\ Rodriguez$^2$}
\author{Laura Sampson$^1$}
\author{Eve Chase$^{1,3}$}
\author{Vassiliki Kalogera$^1$}
\author{Frederic A. Rasio$^1$}
\affiliation{$^1$ Center for Interdisciplinary Exploration and Research in Astrophysics (CIERA) and Dept. of Physics and Astronomy, Northwestern University, 2145 Sheridan Road, Evanston, IL 60208, USA}
\affiliation{$^2$ MIT-Kavli Institute for Astrophysics and Space Research, 77 Massachusetts Avenue, 37-664H, Cambridge, MA 02139, USA}
\affiliation{$^3$ LSSTC Data Science Fellow}

\begin{abstract}
Gravitational waves (GWs) from binary black hole (BBH) mergers provide a new probe of massive-star evolution and the formation channels of binary compact objects. By coupling the growing sample of BBH systems with population synthesis models, we can begin to constrain the parameters of such models and glean unprecedented knowledge about the inherent physical processes that underpin binary stellar evolution. In this study, we apply a hierarchical Bayesian model to mass measurements from a synthetic GW sample to constrain the physical prescriptions in population models and the relative fraction of systems generated from various channels. We employ population models of two canonical formation scenarios in our analysis --- isolated binary evolution involving a common-envelope phase and dynamical formation within globular clusters --- with model variations for different black hole natal kick prescriptions. We show that solely with chirp mass measurements, it is possible to constrain natal kick prescriptions and the relative fraction of systems originating from each formation channel with $\mathcal{O}(100)$ of confident detections. This framework can be extended to include additional formation scenarios, model parameters, and measured properties of the compact binary. 
\end{abstract}

\keywords{gravitational waves, black holes, data analysis, hierarchical Bayesian modeling, stellar evolution}

\section{Introduction}\label{sec:intro}

Recent observations of gravitational waves (GWs) have launched a new branch of observational astronomy. The confident detections of binary black hole (BBH) mergers GW150914, GW151226, and GW170104, as well as BBH candidate LVT151012 by the advanced LIGO (aLIGO) detectors marked the discovery of BBH systems in our universe, and enticed deeper exploration of massive-star evolution \citep{advLIGO,O1BBH,GW150914,GW151226,GW170104}. The final stages of BBH evolution enable the measurement of their physical properties, which connects us to their preceding history and can potentially constrain the environments responsible for facilitating BBH formation, the relative fraction of systems produced through various formation channels, and the physical processes underlying binary stellar evolution. With expected BBH merger rates ranging from 12 to 213 Gpc$^{-3}$yr$^{-1}$ \citep{rates,GW170104}, it is an opportune time to develop methods for utilizing all existing and future observations to constrain and inform astrophysical models.

Two canonical formation channels are generally considered for contributing to the full population of BBHs: isolated binary evolution (i.e., ``the field'') and dynamical formation (i.e., ``clusters''). In the isolated evolution scenario, binaries are predicted to evolve and tighten through a common-envelope phase \citep[e.g., ][]{voss_field,dominik_field,belc_field}, or through chemically homogeneous evolution of close binaries that attain rapid rotation \citep{deMink_chem,marchant_chem,mandel_chem}. Alternatively, the dynamical channel predicts BBHs that become bound and tighten through three-body encounters in dense star clusters such as globular clusters \citep[e.g., ][]{zwart_dyn,downing_dyn1,downing_dyn2,rodriguez_dyn1,rodriguez_dyn2}, galactic nuclei and AGN disks \citep{antonini_nuclear,bartos_AGN,stone_AGN}, or young stellar clusters \citep{ziosi_young,sourav_young}. In addition to these canonical scenarios, more exotic formation channels have been suggested for facilitating BBH mergers, such as field triples tightened by Lidov--Kozai cycles \citep{antonini_triples,silsbee_triples}, primordially formed black holes \citep{pbh}, or remnants of population III stars \citep{popIII}. While these models maintain the ability to predict heavy BBHs such as GW150914 \citep[e.g., ][]{field_GW150914,cluster_GW150194,stevenson_2017b}, the rates and property distributions predicted from population synthesis simulations are highly sensitive to the prescriptions chosen for uncertain physical processes such as black hole natal kick prescriptions, wind mass-loss, and common envelope physics \citep{dominik_CEE,stevenson_2015}.

The analysis of compact binary populations through GW observations provides a unique and powerful mechanism for determining the models that describe the true underlying BBH population. By pairing measured BBH properties from the growing sample of GW observations with population synthesis models that account for various formation scenarios and physical prescriptions, constraints can be placed on the relative fraction of systems produced by each formation channel (i.e., ``branching ratio'') and the inherent physical processes that underpin binary stellar evolution. As the merger rates predicted by various channels are highly uncertain and overlapping, this approach to astrophysical model selection in the context of GWs primarily explores the distributions of the BBH source parameters. Neglecting eccentricity and finite size effects, the masses and spins of component black holes primarily determine the GW signal during the evolution of the BBH through inspiral, merger, and ringdown \citep{properties}, allowing for parameter estimation of these quantities by comparing a measured GW signal with template waveforms generated from a sample of the physical parameters \citep{cutler_PE,poisson_PE,veitch_PE}. 

One confounding aspect of this model selection problem is the uncertain physical mechanisms underlying population modeling of binary compact objects. To circumvent the intricacies and uncertainties of binary evolution modeling, studies such as \cite{mandel_2017} have taken an agnostic, model-independent approach toward model selection by developing methods for distinguishing populations through clustering of source parameters, such as black hole masses. Though such an approach can help to identify multiple populations, it lacks the ability to directly identify the physical processes inherent to stellar evolution models. To this end, \cite{stevenson_2015} assessed the potential of using GW observations for differentiating population synthesis models that have various prescriptions for common envelope binding energy, maximum neutron star mass, black hole natal kick prescriptions, and stellar winds, finding that certain models could be ruled out in the near-future given expected merger rates. However, the inclusion of alternative formation channels would complicate this process. 

Several studies, such as \cite{vitale_2017} and \cite{stevenson_2017}, have performed model selection to infer branching ratios using BBH spin distributions, finding that one may converge upon the branching ratio between field and cluster formation channels with dozens to hundreds of detections. Though spin distributions for differing population models can be constructed purely by geometrical arguments, isolated binary evolution and dynamical formation rely on vastly different procedures for compact binary formation and evolution, and physically motivated modeling is therefore required for accurate and comparable distributions in mass and redshift. Using such models, \cite{spin_tilts} found that certain combinations of masses and spins can be produced exclusively by dynamical formation channels, and the detection of such an outlier could be a clear indication of this formation process. Furthermore, \cite{farr_spins} demonstrated that the clustering of effective spin measurements in the current catalog of GW observations hints at an isotropic spin angle distribution rather than an aligned one. However, there is still much work to be done in utilizing population synthesis and catalogs of GW observations to infer properties of the true compact binary populations. 

In this paper, we present an approach of hierarchical model selection that utilizes physically motivated models of BBH populations from multiple environments to infer underlying physical prescriptions and branching ratios between formation channels. In particular, we focus on the utility of chirp mass measurements for inferring black hole natal kick prescriptions and the branching ratio between isolated binary evolution and dynamical formation models. However, this approach can easily scale to include more measured BBH properties, additional population models, and submodels accounting for different uncertain physical prescriptions.

The outline for the paper is as follows. In Section \ref{sec:bbh_pops}, we discuss the BBH population models used in this analysis, which model field binaries (Section \ref{sub:field}) and cluster binaries (Section \ref{sub:cluster}), accounting for selection biases (Section \ref{sub:detectability}). Section \ref{sec:model_sel} outlines the algorithm for hierarchical model selection (Section \ref{sub:hierarchical}), the mock observations and analytical approximations for measurement uncertainty (Section \ref{sub:mock_obs}), and sampling procedure (Section \ref{sub:RJMCMC}). In Section \ref{sec:inference}, we discuss our inference on branching ratio (Section \ref{sub:beta}) and kick prescription (Section \ref{sub:kicks}) through this methodology. 
We conclude in Section \ref{sec:conclusions} with a discussion of the analysis and future prospects.

\section{BBH Population Models}\label{sec:bbh_pops}

In this section, we describe the population models used in our analysis. These models are identical to those used in \cite{spin_tilts}, except that the various natal kick prescriptions used as submodels of the field population are also incorporated into the globular cluster models. Our inference relies on output parameter distributions of BBHs that merge as potential LIGO sources, such as component masses, spin-tilts, and redshifts. 

\begin{figure*}[t]
\includegraphics[width=\textwidth,height=12cm]{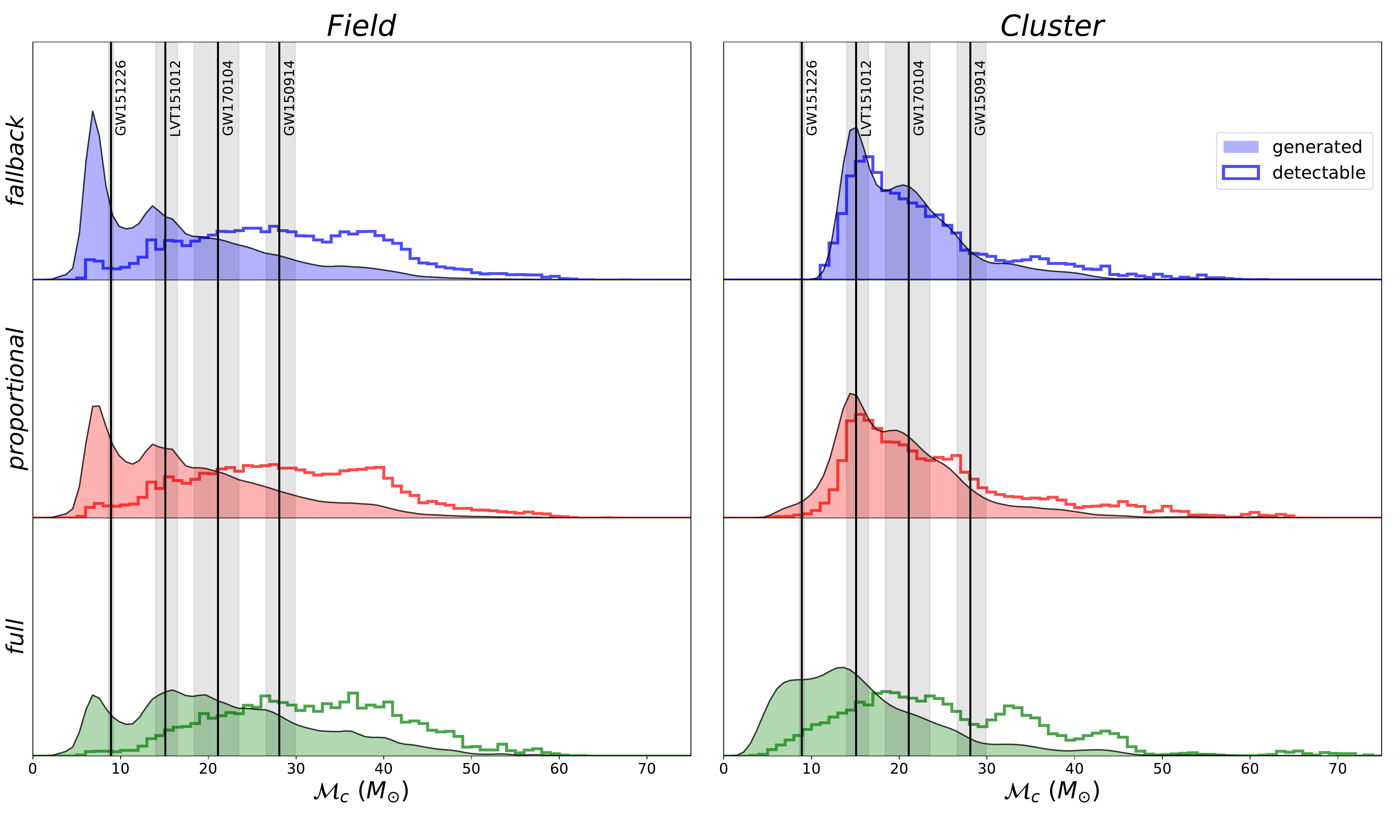}
\caption{Chirp mass distributions for the field and cluster population models. Each panel shows the independently normalized distributions of sources generated (filled histogram) and sources weighted by detectability (unfilled histogram). For reference, the chirp masses of the four likely gravitational-wave events (GW150914, GW151226, GW170104, and LVT151012) are plotted, with the outer lines representing the 90\% credible region. The top, middle, and bottom panels show the distributions for fallback, proportional, and full natal kick prescriptions, respectively. We construct each model using one kick magnitude prescription, comprised of equal abundances from the four submodels described in items \ref{item2} and \ref{item3} of Section \ref{sub:field}. }
\label{fig:chirp_mass_dist}
\end{figure*}

\subsection{Isolated Binary Populations}\label{sub:field}

Populations of field binaries are generated using an upgraded version of the binary evolution code BSE \citep{BSE}. This code rapidly models stellar populations using metallicity-dependent fits for single-star evolution, while also modeling binary interactions such as mass transfer and changes in orbital angular momentum from black hole natal kicks. Furthermore, BSE now uses a radius-dependent fitting formula for the common envelope binding energy parameter $\lambda$ \citep{variable_lambda}. Our modifications to this code implement physical prescriptions from more recent work related to stellar winds and supernova prescriptions. These include mass-loss prescriptions for O and B stars known as the ``Vink prescription'' \citep{vink_winds}, metallicity dependence to the evolution of Wolf-Rayet stars \citep{vink_WR}, and prescriptions for the supernova mechanism developed in \cite{fryer_2012}. Despite these upgrades, many of the physical mechanisms governing binary stellar evolution are still poorly constrained, and incorrect physical prescriptions may propagate inaccuracies to the physical parameter distributions of our populations. We evolve $10^5$ binaries in 11 metallicity bins ranging from $0.005$ to $1.5Z_{\odot}$, with masses from $18M_{\odot}-150M_{\odot}$ drawn from an initial mass function of $p(m)dm \propto m^{-2.3}dm$ \citep[see, e.g.][]{kroupa}. Mass ratios are drawn from a uniform distribution on the interval $[0,1]$, and initial semi-major axes are drawn from a distribution flat in log space on the interval $10R_{\odot}-10^5R_{\odot}$. All binaries that evolve into BBHs are maintained as GW candidates, scaled appropriately by their merger time. 

While the following scenarios may not cover all the physical uncertainties in population models, they do provide a representative sample of possibilities. As such, we consider 12 permutations of physical assumptions that all affect the final population parameters: 

\begin{enumerate}

\item Three different natal kick prescription models, imparting different amounts of linear velocity to the newly formed black holes in the binary. One model, the \textit{fallback} kick prescription, assumes that some fraction of the mass ejected during core collapse will ``fallback" on the black hole: 
\begin{equation}\label{fallback}
V^{\text{BH}}_{\text{kick}}=(1-f_{\text{fallback}})V^{\text{NS}}_{\text{kick}}.
\end{equation}
The fraction of material that falls back is proportional to the core mass of the black hole progenitor. The second model, the \textit{proportional} kick prescription, assumes that the kick imparted to the black hole is reduced by the ratio of the neutron star mass to the black hole mass:
\begin{equation}\label{proportional}
V^{\text{BH}}_{\text{kick}}=\frac{m_{\text{NS}}}{m_{\text{BH}}}V^{\text{NS}}_{\text{kick}}
\end{equation}
where we assume $m_{NS} = 2.5 M_{\odot}$ for all systems, as this value represents the hypothetical `boundary' between neutron stars and black holes in most population synthesis codes. The final kick prescription, called the \textit{full} kick prescription, assumes that the black hole kick is equal to the full kick velocity imparted on the neutron star:
\begin{equation}\label{full}
V^{\text{BH}}_{\text{kick}}=V^{\text{NS}}_{\text{kick}}.
\end{equation}

\item Two differing kick directions. In one model we assume kicks are isotropically distributed in solid angle around the exploding star, which is the common assumption in population models. However, observations of pulsars have suggested a correlation between the kick direction and spin axis \citep{polar_kicks}, motivating the inclusion of a polar kick prescription where the kicks are confined to 10$^{\circ}$ cones about the rotational axis of the progenitor star. \label{item2}

\item Two different methods of accounting for uncertainties in the realignment of the component spin axes after the first supernova. One model allows for realignment of the binary after the first kick, 
whereas the other model does not realign. 
Though this does not have an effect on the mass distributions of the field population models, it has a substantial effect on the spin distributions of the resultant BBHs. \label{item3}
\end{enumerate}

All these variations in model assumptions largely affect the resultant spin-tilt distributions of the binaries. However, only kick magnitudes play a substantial role in the final distribution of BBH chirp masses. As seen in Figure \ref{fig:chirp_mass_dist}, stronger kick prescriptions flatten out the relative abundance of low-mass binaries in field models; these systems acquire larger linear velocities from the kicks, allowing the kinetic energy of the binary component to more easily overcome the gravitational potential and become unbound. As this paper focuses on chirp mass measurements, we construct each model using one kick magnitude prescription, and equal abundances from the four submodels that are described in items \ref{item2} and \ref{item3} of Section \ref{sub:field}. Furthermore, we expect only one kick magnitude to be true, whereas kick direction and binary realignment prescriptions may be dependent on processes such as stellar rotation. Future work will incorporate spin measurements in the inference and address these submodels with more detail.

\subsection{Cluster Binary Populations}\label{sub:cluster}

In this study, we consider the ``classical'' channel of dynamical formation in old, metal-poor globular clusters. Cluster binaries are drawn from a few dozen globular cluster models generated using the Cluster Monte Carlo (CMC) code \citep[see, e.g.][]{CMC}. Black holes sink to the centers of globular clusters due to dynamical friction, fostering mass segregation through energy equipartition and turning the globular cluster cores into dynamical factories for forming heavy stellar-mass BBHs. Though these models are sensitive to initial conditions, dynamically formed binaries rely on single-star evolution and $N$-body dynamics rather than binary stellar evolution and therefore maintain fewer uncertainties in the physical processes (e.g., common envelope evolution, mass transfer) involved in generating BBHs \citep{rodriguez_dyn1}. However, as the choice of kick magnitude prescription may alter the distribution of black holes within a cluster or eject black holes entirely, we include cluster submodels with the same variations in kick prescription presented for the field models in Section \ref{sub:field}. This ensures a putative population mixing has a consistent kick prescription. As seen in Figure \ref{fig:chirp_mass_dist}, the stronger kick magnitude prescriptions tend to flatten out peaks in the normalized chirp mass distributions for the cluster population as well as the field population. Note that in the cluster case the natal kicks do not act to disrupt individual binary systems, as the final partners are usually found long after the components have evolved into compact objects. 

The globular cluster models generated in \cite{rodriguez_dyn1} used the fallback prescription for black hole natal kicks described above. To create equivalent populations using the proportional and full kick prescriptions, we implement the following approximate procedure: for each of the 48 globular cluster models from \cite{rodriguez_dyn1}, we take the initial ($t=0$) snapshot of the cluster, and evolve the massive stars (above $18M_{\odot}$) forward with BSE until the stars have completed their evolution and the initial population of black holes has formed. We then record the velocity of the natal kick, and only retain those black holes for which $V_{\rm{kick}} < \sqrt{-2\Phi(r)}$, where $\Phi$ is the gravitational potential of the cluster, and $r$ is the initial radial position of the star in the cluster.\footnote{Because the black hole-formation timescale for massive stars ($\sim 5$Myr) is significantly smaller than the mass-segregation timescale \cite[$\sim$100 Myr, see][]{oleary_2006}, we can safely ignore the change in position of the star between birth and black hole-formation.}

Once we have an initial population of black holes for each cluster model, we proceed to eject black holes and BBHs from our synthetic population, assuming that the rate at which black holes are ejected is identical to that found with CMC.\footnote{This assumption is well justified, as it is the total energy flux of the cluster, not of the black hole sub-system, that determines the ejection rate of black holes from a globular cluster \citep{breen_heggie_2013}. However, for clusters where there are not sufficiently many black holes to meet the energy requirements of the cluster (e.g., the full kick prescription), this assumption will overestimate the early ejection rate of black holes.} We select black hole masses without replacement from a list of $N$ black holes, according to 

\begin{equation}
p(i)di \propto ~i^2 di~~\rm{for}~ 1 < i < \mathit{N}/3
\label{eqn:cheapHack}
\end{equation}

\noindent where $i$ is the index of the list of black holes, sorted in order of decreasing mass. Equation \eqref{eqn:cheapHack} is physically motivated by the fact that globular clusters preferentially eject the most massive black holes first, continuing to eject black holes until depletion \citep{morscher_2015}. The functional form of $p(i)$ was found through trial-and-error to reproduce the masses and mass ratios of ejected BBHs from the CMC simulations. For every BBH, we also remove four single black holes from the list \citep[see, e.g.][]{heggie_hut}. The reasoning for this is that once a binary is nearing the hardness necessary to eject it from the cluster, the scatterings it undergoes will eject single objects with the same strong three-body encounters responsible for hardening and ejecting the BBH. However, since the binary is about twice as massive as the single objects that are scattered, there are a few scatterings where the velocity of the single object surpasses the cluster escape speed, while the velocity of the binary does not. Numerical tests indicate that on average 3--4 single black holes are ejected for each binary that is ejected. Finally, at the time of ejection, we set the eccentricity and semi-major axis of the binary using the half-mass radius and mass of the cluster at the time of ejection, according to:

\begin{flalign}
&P(e)\,de = 2e\,de\label{eqn:eccen}
\end{flalign}
\begin{flalign}
&P(a | M_{GC}, R_h, \mu_{\text{bin}})\,da = \frac{1}{a \sigma \sqrt{2\pi}} \label{eqn:pa}\times \\ &~~~~~~~~~~~~~~~~~~~~~~~~~~~~~\exp\left[ -\frac{\left(\log \frac{\mu_{\text{bin}} R_h}{a M_{GC}} - a^*\right)^2} {2\sigma^2} \right]\,da \nonumber
\end{flalign}

\noindent where $\mu_{bin} = (m_1 m_2)/(m_1+m_2)$ is the reduced mass of the binary, and $a^{*}$ and $\sigma$ are the parameters of a log-normal distribution with mean $a^{*} = 3.98$ and $\sigma = 0.59$ \citep[see][Equations (7) and (8)]{rodriguez_dyn1}.  The merger time of each binary is computed by adding the time each binary is ejected (assuming all globular clusters to be 12 Gyr old) to the GW merger time from \cite{peters_1964}.  The redshift assigned to each merger is the redshift at that cosmological lookback time, assuming a flat $\Lambda$CDM cosmology with $\Omega_M = 0.306$ and $H_0 = 67.9\text{km}\,\text{s}^{-1}\,\text{Mpc}^{-1}$ \citep{planck_2015}. 

Once we have a population of ejected BBHs from clusters with different kick prescriptions, we resample the population of BBHs to better represent what we expect to see from globular clusters in the local universe. First, we draw a population of binaries from our effective globular cluster models by preferentially selecting binaries from globular clusters closer to the peak of the observed globular cluster mass function \cite[i.e., massive globular cluster models that more-closely resemble the population of observed globular clusters in the local universe; see][]{rodriguez_dyn1}. We then take this population of BBHs from globular clusters, where we have assumed a universal globular cluster mass function and constant spatial density of globular clusters, and create a 3D kernel-density estimate (KDE) of the binary mergers in $m_1$, $m_2$, and redshift. We then draw as many binaries as we want from this distribution using an MCMC \citep{emcee}, with the KDE as our likelihood and a prior on the redshift, which is uniform in comoving volume. 

The stronger kick prescriptions (i.e., full kicks) retain more low-mass binaries relative to their high-mass counterparts. This is because in the full kick case the natal kick velocity does not decrease with increasing black hole mass, contrary to fallback and proportional kicks (see equations \ref{fallback}-\ref{full}. Therefore, the full kick prescription will kick out all black holes from the cluster with equal likelihood regardless of the black hole mass, whereas the velocity of the kick is stifled for higher-mass objects in the fallback and proportional cases, allowing more of these objects to be retained relative to their lower-mass analogs. However, as the stronger kick prescriptions will cause more newly formed black holes to be ejected from the cluster, it also results in a decrease to the overall merger rate. The above procedure is necessary to generate new BBH populations without having to generate new, computationally expensive models of massive globular clusters.  We use this approximate method for all three populations, including the fallback prescription \cite[for which we do have complete CMC models from][]{rodriguez_dyn1}. This was done to avoid any systematic differences that our approximate technique may have introduced, and it was found that this method matched well with the true fallback population generated from CMC.  Finally, we assume that the isotropic and polar kick models for clusters should be identical, since BH retention in clusters should be independent of kick angle.

\subsection{Incorporation of Selection Biases}\label{sub:detectability}

The distributions described above represent all BBH systems that are generated by these populations in the local universe. As the detectability of a given source is dependent on both its physical and orientation parameters (e.g., masses, spins, redshift, frequency content, detector network antenna pattern, inclination), the distribution of observed parameters will be different from the true source distribution. Therefore, we translate the raw source distributions into distributions of detectable sources by the expected design-sensitivity power spectrum and antenna pattern of a single detector
assuming isotropic sky location and inclination distributions. A signal-to-noise (S/N) threshold of 8 is applied, defined by 
\begin{equation}
\rho^2= 4 \Re \int \frac{\tilde{h}^{\star}(f)\tilde{h}(f)}{S_n(f)} df
\end{equation}
where $\tilde{h}(f)$ is the gravitational waveform in the frequency domain and $S_n(f)$ is the one-sided power spectral density of the noise.\footnote{For S/N calculations, we assume non-spinning component black holes and utilize the \texttt{IMRPhenomPv2} waveform approximate \citep{IMRPhenomPv2}.} We then calculate the probability of a system with a given mass and redshift passing this threshold and then weight the distributions accordingly.\footnote{Though the cluster models maintain redshift information, our field populations do not. We assume for simplicity that field binaries are distributed uniformly in comoving volume in the local universe and sample redshifts accordingly.} As seen in the right panel of Figure \ref{fig:chirp_mass_dist}, this tends to flatten out the low-mass peaks and amplify the number of higher-mass systems. 

\begin{figure}[b]
\includegraphics[width=0.48\textwidth]{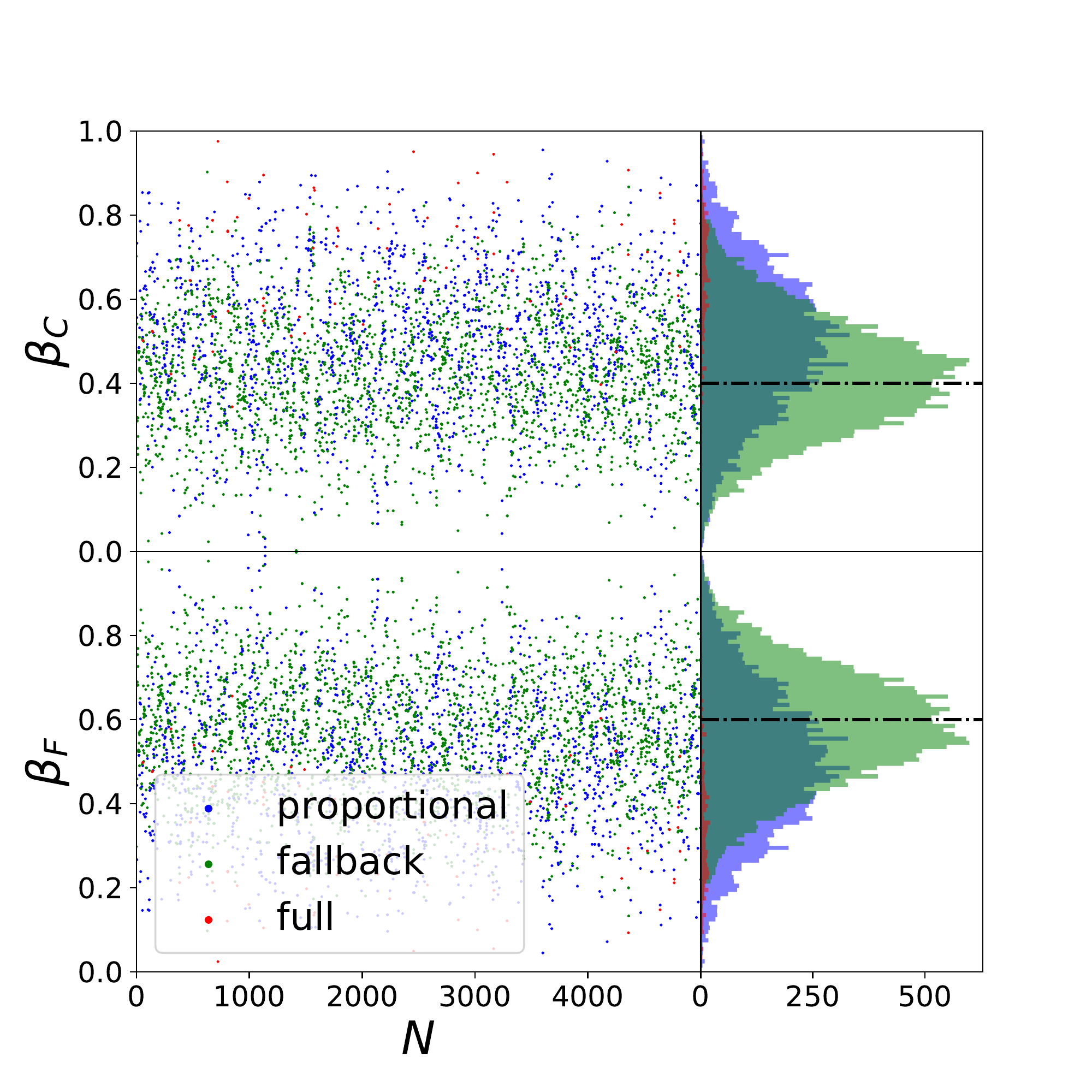}
\caption{Example realization of the sampling, where 5000 samples are drawn from the RJMCMC chain. This particular realization is for 100 observations from the fallback model, with a cluster branching ratio of $\beta=0.4$. $\beta_C$ and $\beta_F$ are the fraction of systems that are drawn from the cluster and field populations, respectively. The left panels show the value of $\beta$ inferred for each step in the sampling, with colors indicating the model chosen. The right panels show the total binned histograms.}
\label{fig:sample_hist}
\end{figure}

\section{Model Selection}\label{sec:model_sel}

With population models in hand, we leverage BBH mass measurements to infer properties of the underlying distribution. The two questions we aim to address in this paper are as follows: 

\begin{enumerate}
\item Given a catalog of $N$ BBH chirp mass measurements from GW observations that come from a population made up of field and cluster binaries with a particular branching ratio ($\beta$), how well can one discern the inherent black hole natal kick prescription?
\item Assuming one prescription is correct, how many observations are required to confidently converge on the true value of this branching ratio?
\end{enumerate}

\noindent We now describe the machinery behind this inference.

\subsection{Hierarchical Modeling}\label{sub:hierarchical}



\begin{figure}[b]
\includegraphics[width=0.48\textwidth]{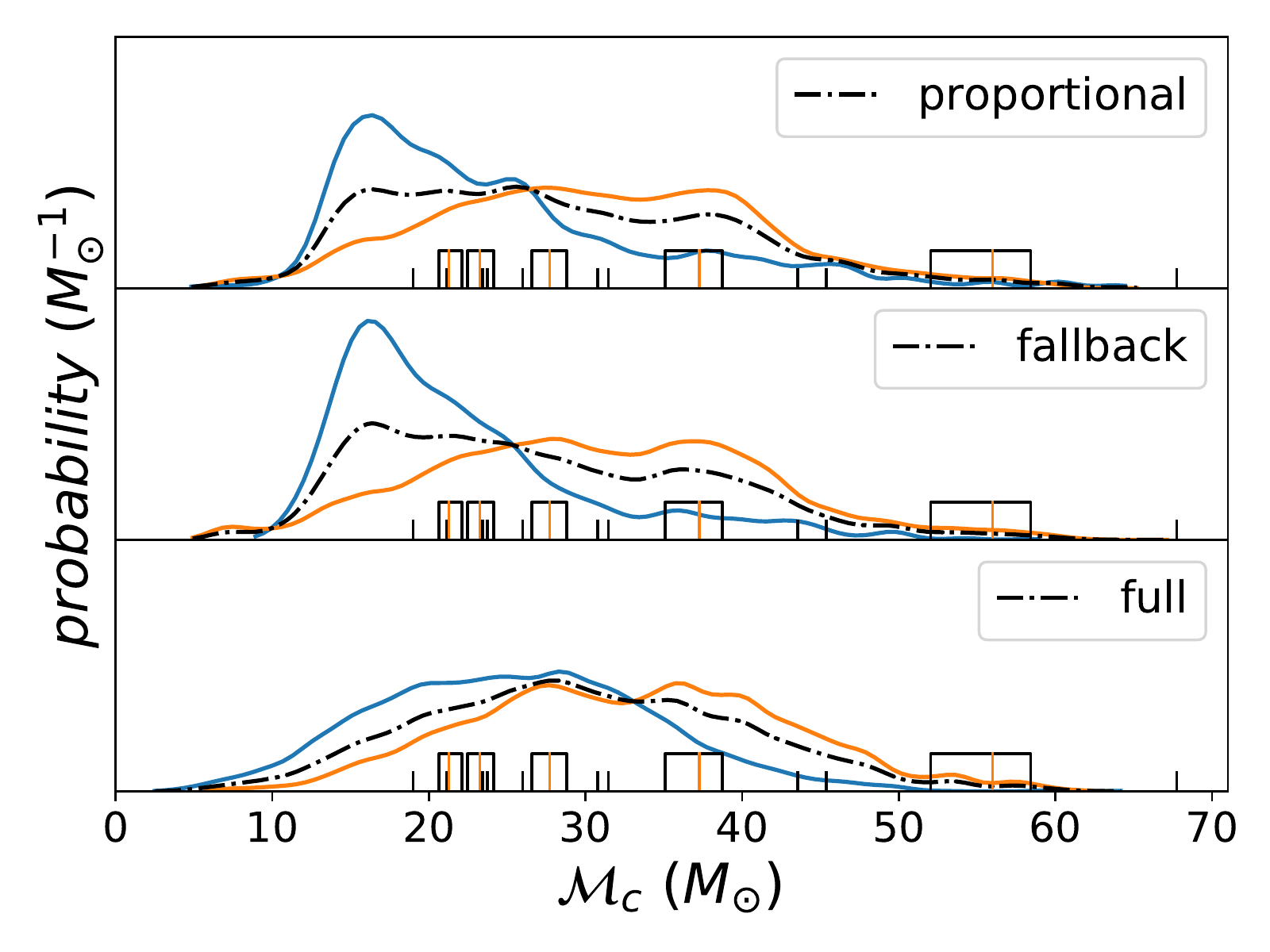}
\caption{Mock observations drawn from a population specified by a kick prescription and branching ratio. This particular realization draws five observations from the fallback kick population model with $\beta = 0.4$. Each panel shows the distributions for a given kick prescription, where the blue and orange lines represent the field and cluster models, respectively, and the dashed black line shows the (normalized) combined population given the value of $\beta$. The box-and-whisker plots at the bottom of each panel show the median value of the posterior samples for an observation with an orange line, the upper and lower quartile of these samples as the edges of the box, and the maximum and minimum value of the posterior samples as whiskers. In this case, all observations are drawn from the fallback population model (dotted line in the middle panel).}
\label{fig:model_obs}
\end{figure}

\begin{figure*}[t]
\includegraphics[width=\textwidth]{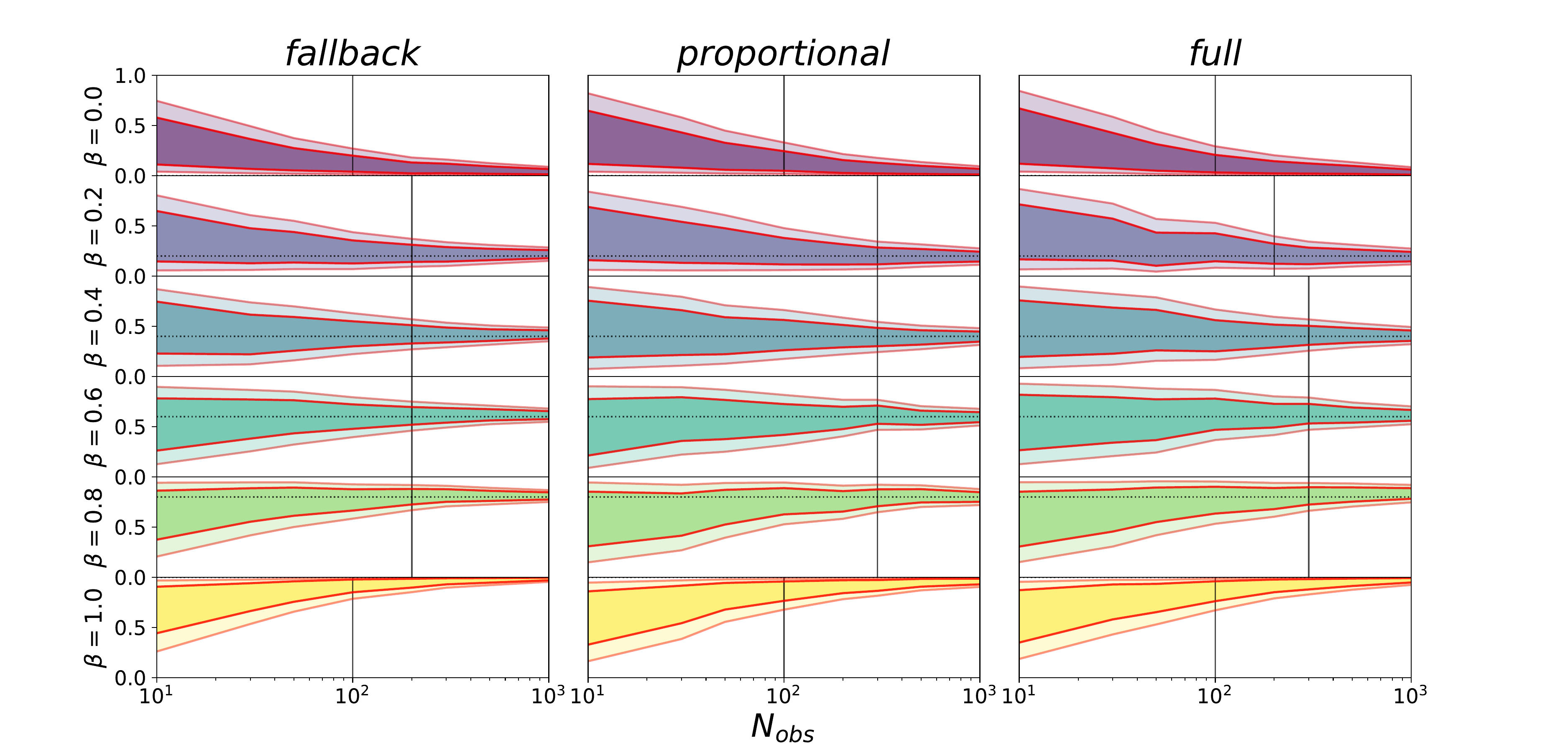}
\caption{Convergence on the true value of $\beta$ as a function of number of observations. The dark and light shadings represent the 68\% and 90\% credible intervals, respectively
. The black vertical lines show the point in our discrete samplings of $N_{\text{obs}}$ at which the 68\% credible interval for $\beta$ is constrained to less than 20\% of the full range of $\beta$; due to our discrete sampling this threshold is in fact reached before this point. The left, center, and right panels show the convergence on $\beta$ when the injected kick prescription is fallback, proportional, and full, respectively. Note that the convergence rate varies depending both on the injected value of $\beta$ and the injected kick prescription.}
\label{fig:beta_convergence_credible_all}
\end{figure*}

As described by \cite{mandel_model_selection} and \cite{hogg_model_selection}, among many others, the objective of hierarchical modeling is to infer a set of model parameters $\vec{\lambda}$ given $N$ observations $\{x_i\}$, which are characterized by a set of physical parameters $\{\vec{\theta}_i\}$ and constrained by prior assumptions $\{\vec{\alpha}_i\}$. The astrophysical model described by parameters $\vec{\lambda}$ gives a probability distribution for physical parameters, in our case chirp masses. By Bayes' theorem, the posterior on $\vec{\lambda}$ is 

\begin{equation}
p(\vec{\lambda}|\vec{\theta})=\frac{p(\vec{\theta}|\vec{\lambda})p(\vec{\lambda})}{p(\vec{\theta})}
\end{equation}

\noindent where $p(\vec{\theta}|\vec{\lambda})$ is the likelihood of observing a particular set of physical parameters, $p(\vec{\lambda})$ is the prior on the model parameters, and $p(\vec{\theta})$ is a normalization constant. 

However, as we observe $N$ independent GW signals rather than the physical parameters directly, we rewrite the likelihood as

\begin{equation}
p(\{x_i\}|\vec{\lambda}) = \prod_{i=1}^N p(x_i|\vec{\lambda}) = \prod_{i=1}^N \int d\vec{\theta}p(x_i|\vec{\theta})p(\vec{\theta}|\vec{\lambda})
\end{equation}

\noindent Again applying Bayes' Theorem, we write $p(x_i|\vec{\theta})$ as $p(\vec{\theta}|x_i)p(x_i)/p(\vec{\theta})$ to get

\begin{equation}
p(\{x_i\}|\vec{\lambda}) = \prod_{i=1}^N p(x_i) \int d\vec{\theta}\ \frac{p(\vec{\theta}|x_i)p(\vec{\theta}|\vec{\lambda})}{p(\vec{\theta})}
\end{equation}

\noindent where $p(\vec{\theta})$ is the prior on the physical parameters that are used to generate the posterior samples.

We approximate the integral as a discrete sum over posterior samples

\begin{equation}
\int d\vec{\theta} p(\vec{\theta}|x_i)f(\vec{\theta}) \approx \frac{1}{S} \sum_{k=1}^S f (\vec{\theta}_k)
\end{equation}

\noindent and ignore the multiplicative constant $p(x_i)$ as it is not a function of $\vec{\theta}$ or $\vec{\lambda}$ and will not affect the sampling of the posterior. Therefore, the full expression for the likelihood that we wish to sample is 

\begin{equation}
p(\{x_i\}|\vec{\lambda}) = \prod_{i=1}^N \frac{1}{S}\sum_{k=1}^S \frac{p(\vec{\theta}_k|\vec{\lambda})}{p(\vec{\theta}_k)}
\end{equation}

\noindent where, again, $N$ is the number of observed events, $S$ is the number of posterior samples, $\vec{\theta}_k$ are the astrophysical parameters, and $\vec{\lambda}$ are the model parameters.

We aim to do model selection between different kick prescriptions while simultaneously performing inference on branching ratios between field and cluster formation channels. The parameters of our astrophysical model are thus $\vec{\lambda}=(\iota,\beta)$, where $\iota$ is an indexing parameter that indicates the kick prescription ($\iota \in [0,1,2]$ where 0, 1, and 2 designate the proportional, fallback, and full kick prescriptions, respectively) and $\beta$ is the branching ratio parameter, defined as the fraction of observations that are drawn from cluster models ($0 \leq \beta \leq 1$). 

As the branching ratios between the various formation channels are highly uncertain, we maintain minimal assumptions on our prior knowledge of the model parameters. For the prior on $\beta$, we use a Dirichlet distribution, which is a multivariate generalization of the beta distribution. This allows for minimal prior assumptions while ensuring the values of $\beta$ for all channels sum to unity. Though X-ray binary observations \citep{xrb_kicks} and the current catalog of BBH observations \citep{field_GW150914} provide moderate evidence for certain black hole natal kick prescriptions, we use a uniform prior on the kick prescription, which for this discrete parameter puts equal weight on each prescription. 

\subsection{Mock observations}\label{sub:mock_obs}

We represent the chirp mass distributions for our populations with a Gaussian KDE and draw `observations' from this model. In practice, the observations themselves are manifested as a set of samples drawn from a posterior computed for each candidate event. Instead of employing very accurate, but computationally expensive, Markovian methods to estimate the parameter posteriors, we instead use the Fisher matrix as a proxy for the inverse covariance of a simpler Gaussian parameter distribution \citep[see][]{cutler_PE}. We are justified in this procedure in the case of chirp mass ($\mathcal{M}_c$), because it determines the leading-order evolution of GWs from compact binary coalescence and is therefore the best-measured physical property from a GW signal \citep[see][and references therein]{properties}. However, this methodology is less accurate for parameters that are more correlated and less constrained, such as effective spin and symmetric mass ratio \citep[see, e.g.][]{fisher_vallisneri,fisher_bad1,fisher_bad2}.


As the Fisher matrix tends to overestimate the distributional width in $\mathcal{M}_c$, this also provides a conservative estimate for the true measurement uncertainty. However, as chirp mass measurements are highly constrained, we find that the inclusion of measurement uncertainty does not drastically affect our results. Nonetheless, we draw 100 mock posterior samples from a Gaussian distribution with a mean centered on the true value and standard deviation $\sigma_{\mathcal{M}_c}$ as each `observation'. 

$N$ smeared observations are drawn from a `true' distribution, described by a particular natal kick prescription and a value of the branching ratio $\beta$ (such that $N_{cluster} = \beta N$ and $N_{field} = (1-\beta)N$). Figure \ref{fig:model_obs} shows one realization of this procedure. We then use these mock observations as the basis behind our statistical inference through hierarchical modeling.

\subsection{Sampling}\label{sub:RJMCMC}

\begin{figure*}[p]
\includegraphics[width=0.98\textwidth, height=23cm]
{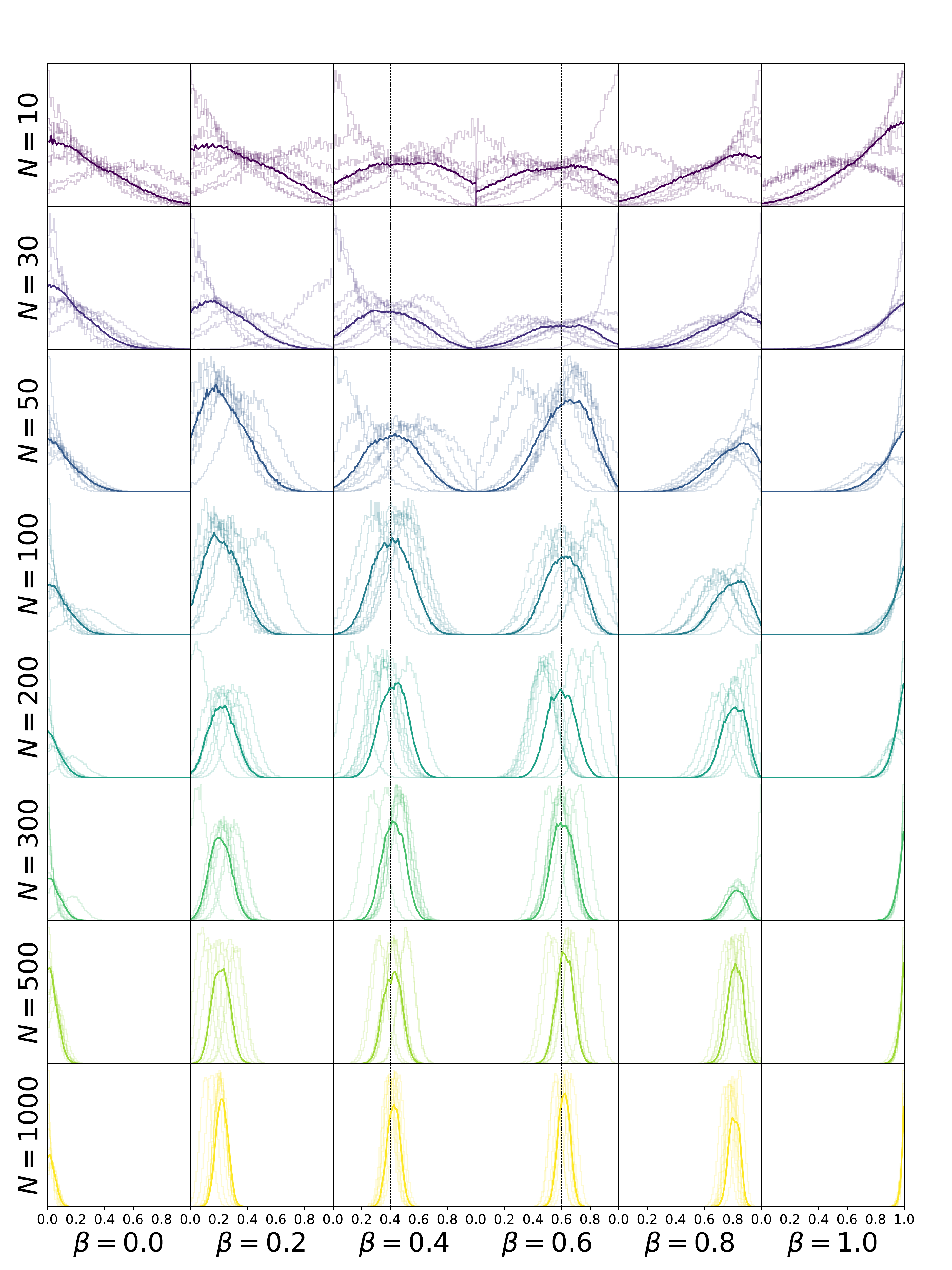}
\caption{Convergence on the branching ratio $\beta$ for various number of observations and injected values of $\beta$. Colors represent the number of observations, and are identical to the colors used in Figure~\ref{fig:beta_convergence_credible_all}. All observations are drawn from a population in which the `fallback' model is deemed true. In each panel, the 10 realizations of the inference on $\beta$ are shown as faded lines, and the dark line is the median value of 100 realizations at each bin of the histogram. The dashed black vertical line marks the injected value of $\beta$.}
\label{fig:beta_convergence}
\end{figure*}

The technique we use for sampling the posterior on the model parameters is Reverse Jump Markov Chain Monte Carlo \citep[RJMCMC; see e.g.][]{rjmcmc}. In this method, the calculation of the Bayes factor between two models does not require the explicit calculation of the evidence integral. Rather, the model itself becomes a parameter of the chain. Depending on which value an indexing parameter takes, the likelihood and prior are evaluated using one of a set of models, which may or may not be of the same dimensionality. Said another way, the sampler first `jumps' in model index space, and then estimates the value for $\beta$ within the particular model it lands. We assume that all models contain the same set of model parameters with the same meaning --- in other words, the branching ratio $\beta$ is the fraction between field and cluster in all models. The samples can then be sorted by model index at the end in order to generate posteriors for the individual kick prescriptions. Since our models have the same number of parameters with the same priors, the Bayes factor is then simply the ratio
of the number of iterations that the chain spends in each model: 

\begin{equation}
B_{ij} =  \frac{\text{\#\ of\ iterations\ in\ model\ i}}{\text{\#\ of\ iterations\ in\ model\ j}}
\end{equation}

\noindent We use the open source MCMC python library \texttt{emcee} \citep{emcee} for the implementation of this algorithm.

\section{Astrophysical Inference with Chirp Mass Measurements}\label{sec:inference}

We now seek to utilize this inference for constraining branching ratios and properties of stellar evolution, given a catalog of BBH observations. Because the current number of BBH observations is likely too few to make any substantial claim about formation channels or physical prescriptions, we demonstrate the method using a mock catalog of BBH observations. We then apply one kick prescription model and formation channel branching ratio value to be the `correct' description of nature, and gauge how well one can converge on these injected values over an increasing number of observed systems. For the purposes of this study, we consider 100 realizations for each combination of the branching ratio $\beta$, kick prescription $\iota$, and number of observations $N_{\text{obs}}$.

\subsection{Branching Ratios}\label{sub:beta}

\begin{figure}[b]
\includegraphics[width=0.5\textwidth]{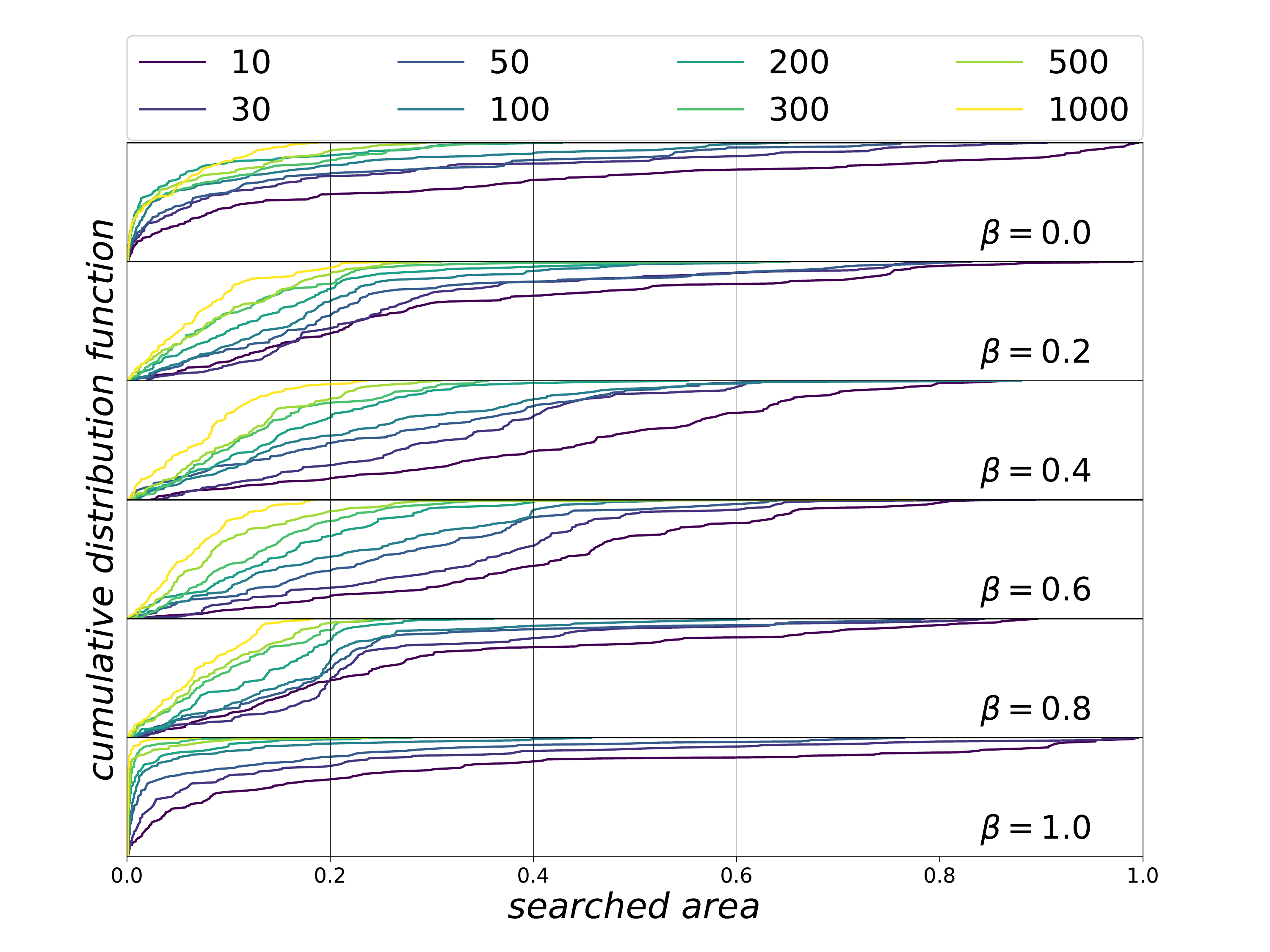}
\caption{Cumulative distribution function on searched area for various injected values of $\beta$ and number of observations (colored lines). The searched area represents the distance between the mode of the distribution and its true value. The searched area for significant $N_{\text{obs}}$ is notably smaller when one channel dominates, e.g. $\beta \sim 1 \text{ or } \beta \sim 0$.}
\label{fig:searched_area}
\end{figure}

Convergence to the true branching ratio is a strong function of the number of observations drawn from the true population, and is also sensitive to the injected branching ratio itself. To summarize the convergence as a function of observations, we plot the marginalized posterior on $\beta$ for different values of $\beta$ in Figure \ref{fig:beta_convergence_credible_all}. To demonstrate this convergence in another way, Figure \ref{fig:beta_convergence} shows the injected value of $\beta$ for different combinations of $\beta$ and $N_{\text{obs}}$, (i.e., many realizations of the sampling visualized in Figure \ref{fig:sample_hist}), as well as the median value of the 100 realizations for each combination. As expected, there is rapid convergence on the true value of $\beta$ as the number of observations increases, as well as smaller variance in the individual realizations. From Figure \ref{fig:beta_convergence}, it is also noted that the inference on $\beta$ is unbiased. 

Other model parameters aside, with only chirp mass measurements we converge on the true branching ratio to an accuracy of $\pm$10\% with $\mathcal{O}(100)$ observations of BBH systems. This convergence on branching ratio is similar to that found in \cite{vitale_2017} and \cite{stevenson_2017} using population models with varied spin distributions; with dozens to a hundred observations, we will begin to see strong convergence on the true value of the branching ratio if BBH observations are dominated by the two canonical channels. Cumulative distribution functions of the searched intervals for $\beta$ are shown in Figure \ref{fig:searched_area}. When only one mode in the distribution is present, the searched interval refers to the distance between the mode of the distribution and its true value. Quantitatively, when one channel dominates the overall event rate, convergence on the true value is noticeably better.


\begin{figure}[t]
\includegraphics[width=0.47\textwidth]{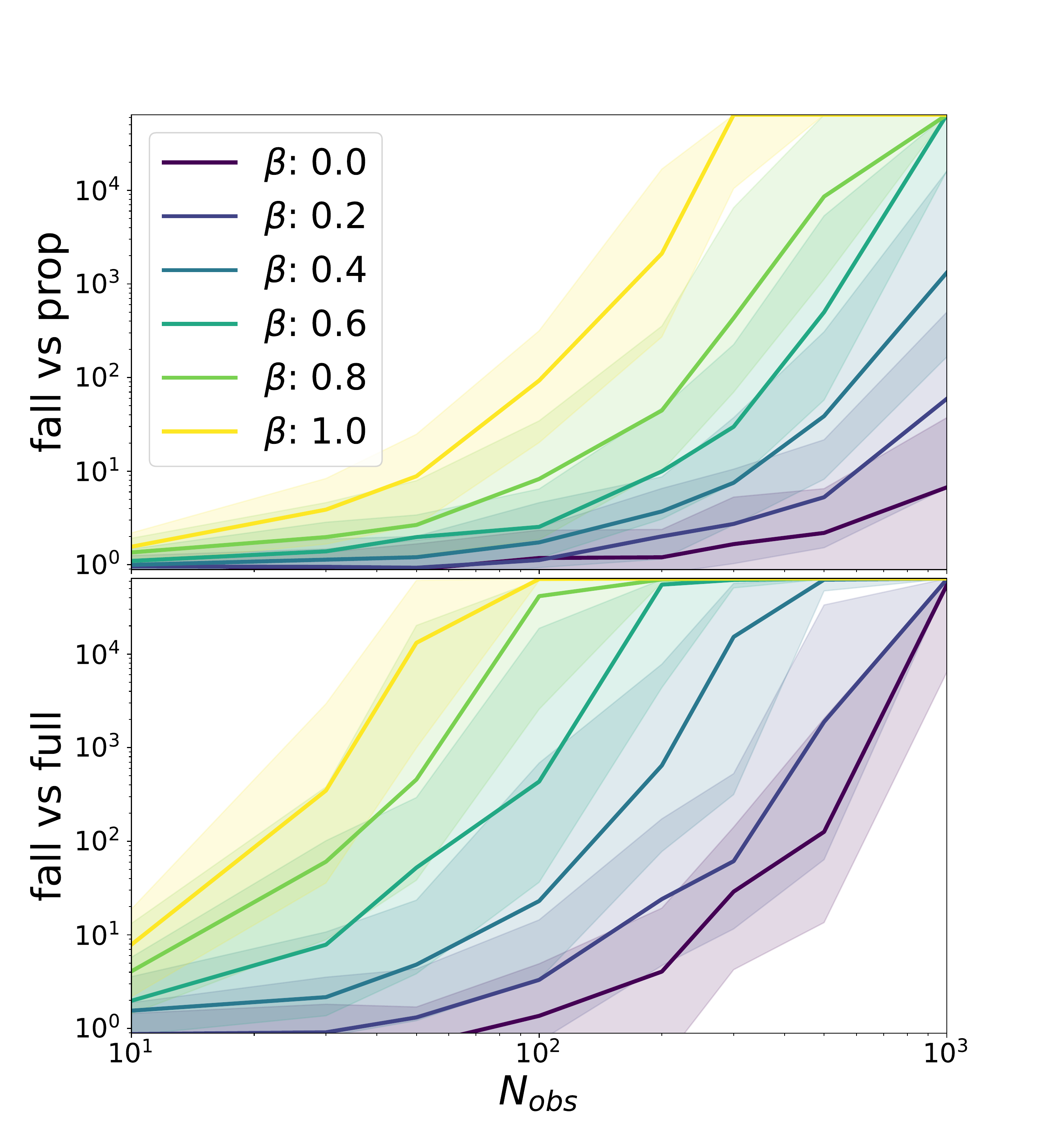}
\caption{Bayes factors between different models as a function of the number of observations. The dark lines show the median value, and shaded regions a 68\% credible interval derived from the 100 realizations. For this figure, observations were drawn from the `fallback' kick prescription, and various injected values of $\beta$ are shown with different colors. The top and bottom panels show the Bayes factor between fallback kicks and proportional kicks, and the Bayes factor between fallback kicks and full neutron star kicks, respectively. The upper limit of the plot is the maximum Bayes factor calculable given our number of samples, and leads to the apparent flattening of the function near this maximal value. The Bayes factor between fallback and full kick models increases much more rapidly than the Bayes factor between fallback and proportional kick models as a function of $N_{\text{obs}}$, since the chirp mass distributions produced by fallback and full kick models are morphologically much more distinct.}
\label{fig:bayes_factor}
\end{figure}

\subsection{Natal Kick Model Selection}\label{sub:kicks}

Our methodology also provides inference on the model index -- that is, the underlying physical prescriptions assumed in the models. For the purposes of this study, the only physical prescription altered between models was the black hole natal kick magnitude. 
Since all models have the same parameters and prior ranges, the Bayes factor for one model compared to another can be simply computed as the ratio of the number of iterations the chain spends in each model. 

As the kick prescription has a noticeably different effect on the distribution of chirp masses in detectable cluster models relative to detectable field models (see Figure \ref{fig:chirp_mass_dist}), the confidence for one prescription relative to another is a strong function of the branching ratio, as well as the number of observations. For example, branching ratios closer to $\beta=1$ draw more observations from the cluster models, which have more distinctive features in the physical parameter distributions of detectable binaries compared to field models and allow for easier discrimination between populations. Furthermore, the growth of Bayes factors as a function of $N_{\text{obs}}$ is expedited when comparing two kick prescriptions with dramatically different effects on the physical parameter distributions. This can also be seen in Figure \ref{fig:bayes_factor}: the Bayes factor between fallback kick and full kick increases much more rapidly than the Bayes factor between fallback kick and proportional kick. 

\begin{figure}[b]
\includegraphics[width=0.47\textwidth]{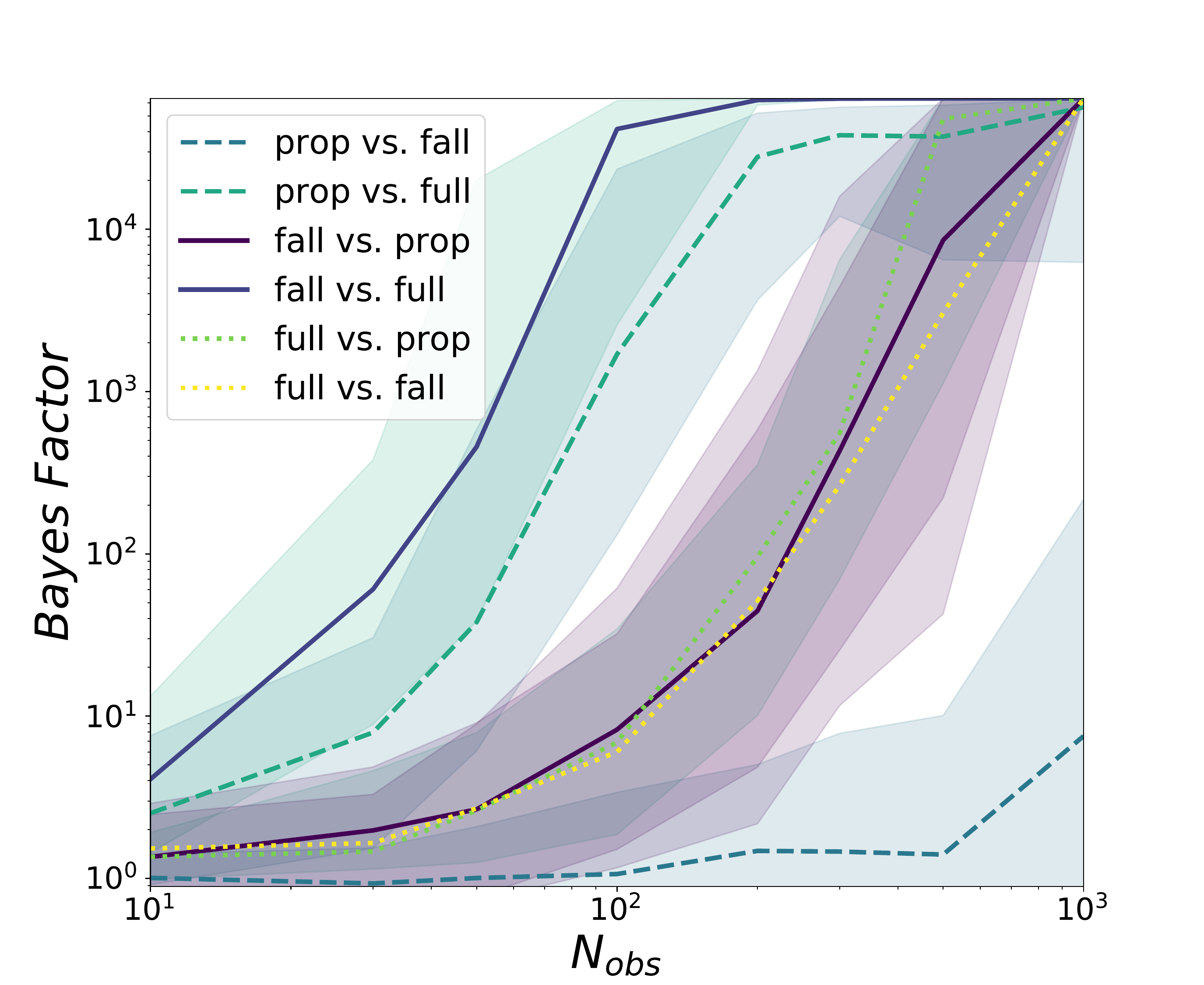}
\caption{Bayes factors between different models as a function of $N_{\text{obs}}$, with observations drawn from models with different injected kick prescriptions. Dotted lines, dashed lines, and solid lines represent full kicks, proportional kicks, and fallback kicks as the correct distribution from which observations are drawn, respectively. All models in this plot have an injected value of $\beta=0.8$. The median values and credible regions are indicated as in Figure \ref{fig:bayes_factor}. Interestingly, we find that when the proportional kick model is injected as the `true' model, our inference does not necessarily prefer proportional kicks as a function of $N_{\text{obs}}$. This is particularly apparent when comparing the Bayes factor between proportional kicks and fallback kicks, when observations are drawn from the proportional kick model. Though the populations are strikingly similar, we believe this issue arises from our conservative approximation of measurement uncertainty. Our approximations, which rely on Fisher matrix formalism for determining the spread of the posterior distribution, provide symmetric widths in our sampled posterior to both lower and higher values for $\mathcal{M}_c$. This may be unrealistically bolstering posterior samples in the low-mass peak of the fallback models (see Figure \ref{fig:model_obs}). Furthermore, as we limit to $\delta$-function observations, this effect disappears.}
\label{fig:bayes_factor_all}
\end{figure}

We achieve Bayes factors between natal kick prescriptions of $\sim$20 as the number of observations reaches $\mathcal{O}(100)$, though, as can be seen in Figure \ref{fig:bayes_factor}, the rate of increase of Bayes factors is extremely sensitive to the injected value of the branching ratio. Given the predicted discovery rates and projected interferometer sensitivity increases in the next few years, this indicates that we can begin to confidently infer the natal kick from supernovae within the lifetime of a design-sensitivity interferometer network. Even if the detection rate remains low, ruling out one physical model compared to others is within reach, especially given the conservative assumptions we have made. As population models are parameterized by many other discrete and continuous variables, we can expand this analysis to constrain other uncertain physical prescriptions of population synthesis using upcoming GW observations. Furthermore, this hierarchical approach does not require the models to be parameterized in the exact same way.









\section{Discussion and Conclusions}\label{sec:conclusions}

With the detection rates predicted for the advanced network of GW observatories, we can look forward to dozens to hundreds of BBH observations in the next decade. These systems provide a unique tool for studying massive-star evolution and the environments in which BBH systems arise, and by pairing a catalog of detections with detailed population models we can begin to constrain many of the uncertain processes driving binary stellar evolution. 

This work investigates how hierarchical modeling can infer the parameters of binary stellar evolution from multiple formation channels using solely chirp mass measurements. We find that with $\mathcal{O}(100)$ observations, we will see convergence on the value for the branching ratio and the preferred natal kick prescription, provided the two channels considered dominate BBH rates. Furthermore, `extreme' values of $\beta$ (i.e., domination by a single formation channel) facilitate quicker convergence on both the branching ratio and natal kick prescription. 

Notably, we find a trade-off between inference on the branching ratio and inference on the natal kick prescription. This effect is dependent on the injected kick prescription itself; models with the fallback prescription shows the largest disparity between field and cluster models, thereby allowing quicker convergence on branching ratio, whereas models using the full kick prescription are less distinguishable and require more observations to converge on the branching ratio, as seen in Figure \ref{fig:beta_convergence_credible_all}. However, as full kicks predict drastically different combined distributions of field and cluster populations relative to fallback and proportional kicks, comparisons with this model allow for accelerated inference on kick prescription, as we demonstrate in Figure \ref{fig:bayes_factor_all}. 

Using our methodology, the current number of GW observations from BBH systems is far fewer than the number of observations needed to make any meaningful statement about kick prescription and branching ratio. Nonetheless, we inject the chirp masses of the three current GW events and one GW candidate from the first and second aLIGO observing runs of as our observations, using the 90\% credible intervals for chirp mass measurements cited in \cite{O1BBH} and \cite{GW170104} to generate mock posterior samples. As expected, our analysis recovers the priors for both quantities and provides no discernment on branching ratio and true natal kick prescription. 

The methodologies in this paper provide a framework for many extensions and refinements, both in the context of inferring additional parameters of population models and including more measured properties of BBH systems. One very simple extension, for example, is to measure the actual physical event rates, and compare with the rates derived from GW observations. Another possibility would be to constrain parameters that define a mass gap, either at low masses, between the maximum observed neutron star mass and minimum observed black hole mass, or at high masses due to pair instability supernovae. 

Multiple studies, such as \cite{stevenson_2017} and \cite{vitale_2017}, have also found that inference using spin distributions can converge on branching ratios with a similar number of GW observations. Though subject to higher measurement uncertainty and highly influenced by an unknown spin magnitude distribution, the inclusion of spin parameters in our analysis could help distinguish models, as dynamical formation predicts an isotropic distribution in spin-tilts whereas isolated field binaries are believed to preserve the memory of their initial spin alignment and can further align their spins through mass transfer, common envelope evolution, and tidal torquing. Furthermore, kick prescriptions have a large effect on spin distributions and the inclusion of spin parameters could act to bolster the confidence of one kick model compared to another. Notably, the detection of a single outlier event, such as those described in \cite{spin_tilts}, could go a long way toward discriminating between models. 

The inclusion of spins would also allow inference on other poorly constrained model parameters of population synthesis; though they have minimal impact on the mass distribution, the direction of the natal kick or partial realignment of the binary after the first supernova, for example, strongly affect the resultant spin distributions in population synthesis models. As the models used in this study are equipped with spin-tilts for the BBH systems, we plan to include spin measurements for the purposes of model selection in future work. However, we note that other mechanisms unaccounted for in our models likely affect the spin-tilts of the binary components as well, and may act to contaminate the information we extract about the physical processes that are accounted for. 


Besides taking a hierarchical approach to model selection using chirp mass distributions, this study provides the framework for determining Bayes factors between a discrete set of population models. In future work, we plan to expand this methodology to include other population synthesis model parameters, such as common envelope efficiency, natal kick direction, and the rate of binary coalescences as a function of redshift. Furthermore, the framework is extensible, allowing the inclusion of other proposed formation channel models, such as young stellar clusters, galactic nuclei, and chemically homogeneous evolution. As more BBHs are observed by the advanced network of GW detectors, this type of inference will evolve into a powerful tool for constraining the correct physical prescriptions in population synthesis models, thereby improving our understanding of the physical processes governing binary stellar evolution. 


\acknowledgments
We would like to thank Simon Stevenson, Steven Reyes, and our anonymous referee for helpful suggestions on this manuscript, as well as Scott Coughlin for assistance in debugging code. This work was supported by NSF Grant AST-1312945, NSF Grant PHY-1607709, and NASA Grant NNX14AP92G at Northwestern University. M.Z. greatly appreciates financial support from the IDEAS Fellowship, a research traineeship program supported by the National Science Foundation under grant DGE-1450006. C.R. is grateful for the hospitality of the Kavli Institute for Theoretical Physics, supported by NSF Grant PHY11-25915, and is supported at MIT by a Pappalardo Fellowship in Physics. L.S. acknowledges support from the L'Oreal FWIS Fellowship program. E.C. thanks the LSSTC Data Science Fellowship Program; her time as a fellow has benefited this work. V.K. and F.A.R. also acknowledge support from NSF Grant PHY-1066293 at the Aspen Center for Physics. The majority of our analysis was performed using the computational resources of the Quest high performance computing facility at Northwestern University, which is jointly supported by the Office of the Provost, the Office for Research, and Northwestern University Information Technology. This paper has been assigned LIGO document number ligo-P1700064.


\bibliographystyle{yahapj}
\bibliography{references}

\end{document}